\documentclass[aps,prd,
reprint]{revtex4-2}
\usepackage[LGR,T1]{fontenc}
\usepackage[latin9]{inputenc}
\usepackage{graphicx}
\usepackage{color}
\usepackage{textcomp}
\usepackage{dsfont}
\usepackage{amsmath}
\usepackage{esint}

\makeatletter

\ProvideTextCommand{\~}{LGR}[1]{\char126#1}

\makeatother

\usepackage{babel}
\begin{document}
\title{{\normalsize{}Supersymmetric quantum mechanics and the 
Riemann hypothesis}}
\author{Pushpa Kalauni$^{a}$ }

\email{pushpakalauni60@gmail.com}
\author{Kimball A Milton$^{b}$}
\email{kmilton@ou.edu}
\date{\today}
\affiliation{$^{a}$Department of Physics, Indian Institute of Space Science and
Technology, Thiruvananthapuram, Kerala, 695547, India~\\
 $\ensuremath{^{b}}$Homer L. Dodge Department of Physics and Astronomy,
University of Oklahoma, Norman, Oklahoma 73019, USA}
\begin{abstract}
{\normalsize{}We construct a supersymmetric quantum mechanical model
in which the energy eigenvalues of the Hamiltonians are the products of Riemann
zeta functions. We show that the trivial and nontrivial zeros of the
Riemann zeta function naturally correspond to the vanishing ground state
energies in this model. The model provides a natural form of supersymmetry.
}{\normalsize\par}
\end{abstract}
\maketitle

\section{Introduction}
\label{sec:1}

In this paper, we propose a supersymmetric quantum
mechanical model in which the eigenvalues of the
Hamiltonian are given in terms of Riemann zeta
functions. Because the model is supersymmetric, the
ground state energy should be zero, which, in fact, leads
to the trivial and nontrivial zeros of the zeta function.
For studying the nontrivial Riemann zeros in the critical strip,
$0<\Re(s)<1$, we have proposed a new approach in a
supersymmetric quantum model \cite{das1}, in which the 
energy eigenvalues of the Hamiltonian are related to the Riemann zeta function
$\zeta(s)$. Furthermore, we have shown that the Riemann zeros on
the critical line ($\Re(s)=1/2$) appear naturally from requiring the vanishing
the ground state energy condition in the model. 

Now, the natural question arises, can it be possible to obtain 
a supersymmetric  model which can provide the trivial and nontrivial zeros
of the zeta function as a vanishing ground state energy? If there
exists a model, then how is the Hilbert space defined? And what are the 
boundary conditions, and how can the orthogonality and completeness of the 
eigenfunctions be demonstrated?
For these reasons, we define a supersymmetric model on a finite interval of
the real line, and impose appropriate self-adjoint boundary conditions.
We verify that the states are complete and orthonormal.
To include trivial zeros of the zeta function in the
supersymmetric system, we introduce a real parameter $\mu$ and define
a modified inner product. Using the modified inner product, we show
that the ground state energy of our supersymmetric model vanishes
for the trivial as well as nontrivial zeros of the zeta function.
For $\mu=0$ this modified inner product reduces to the standard Dirac
inner product, and the ground state energy vanishes for
the discrete nontrivial zeros of the zeta function on the line  $\Re(s)=1/2$. 

\section{Supersymmetric partner Hamiltonians}
\label{sec:1}

 We start with a simple generic description of supersymmetry in one-dimensional
supersymmetric quantum mechanics \cite{witten,cooper,das2}. In supersymmetric
quantum mechanics, the supersymmetric partner Hamiltonians are given
by $H_{-}=A^{\dagger}A$ and $H_{+}=AA^{\dagger}$, where $A$ and
$A^{\dagger}$ are lowering and raising operators. We can write the
Hamiltonian $H$ in the form of a 2 $\times$ 2 matrix as 
\begin{equation}
H=\left(\begin{array}{cc}
H_{-} & 0\\
0 & H_{+}
\end{array}\right)=  \left(\begin{array}{cc}
A^{\dagger}A & 0\\
0 & AA^{\dagger}
\end{array}\right).\label{eq:5}
\end{equation}

The operators $A$ and $A^{\dagger}$ act on the $n^{th}$ state $\psi_{n}$
in the following manner ($n$ is an integer):
%\begin{subequations}
\begin{equation}
A\psi_{n}  =c_{n}\psi_{n-1}, \quad
A^{\dagger}\psi_{n}  =c_{n+1}^{*}\psi_{n+1}.\label{eq:8}
\end{equation}
%\end{subequations}
The supersymmetric partner Hamiltonian $H_{-}$ acts on a state $\psi_{n}$
according to
\begin{equation}
H_{-}\psi_{n}=  A^{\dagger}A\psi_{n}=E_{n}\psi_{n},\quad E_n=|c_{n}|^{2}.
\label{eq:9}
\end{equation}

When $H_{+}$ acts on the partner state $\tilde{\psi}_{n}=A\psi_n$, it satisfies
\begin{equation}
H_{+}\tilde{\psi}_{n}  =E_{n}\tilde{\psi}_{n}.\label{eq:13}
\end{equation}
Equations~(\ref{eq:9}) and (\ref{eq:13}) show that $\psi_{n}$
and $\tilde{\psi}_{n}$ are the supersymmetric partner states with
the same energy $E_{n}>0$, as long as $A\psi_{n}$ does not vanish.
However, the ground state $\psi_{0}$ of the supersymmetric system
is unique and satisfies
\begin{equation}
A\psi_{0}=0,\,\,  H_{-}\psi_{0}=A^{\dagger}A\psi_{0}=0,\,\,E_0=0.\label{eq:14}
\end{equation}
%The supersymmetric partner states $\psi_{n}$ and %$\tilde{\psi}_{n}$
%satisfy
%\begin{subequations}
%\begin{eqnarray}
%\psi_{n}&= & %\frac{(A^{\dagger})^{n}}{c_{1}^{*}c_{2}^{*}..c_{n}%^{*}}\psi_{0},\,\,\,\,\tilde{\psi}_{n}=A\psi_{n},\%,\, \tilde\psi_0=0, \\
%H_{-}\psi_{n} & %=&E_{n}\psi_{n},\,\,\,\,\,\,\,H_{+}\tilde{\psi}_{n%}=E_{n}\tilde{\psi}_{n},\,\,n\ge0.\label{eq:15}
%\end{eqnarray}
%\end{subequations}

Starting in Sec.~\ref{sec:model},  
we present a model of a supersymmetric quantum system, 
in which the energy eigenvalues are given
in terms of the Riemann zeta function, which we review
in Sec.~\ref{sec:2}. 
We show that
the trivial and nontrivial zeros of the Riemann zeta function correspond
to the vanishing ground state energy of the system. 

\section{The Riemann Hypothesis}
\label{sec:2}
In 1859, Riemann \cite{riemann} made a conjecture regarding the nontrivial
zeros of the Riemann zeta function, known as the Riemann hypothesis.
The Riemann hypothesis is directly associated with the understanding
of the distribution of prime numbers, which are the building blocks of
all numbers. The connection between the zeta function and prime
numbers was made by Euler in the following form \cite{edwards,titchmarsh}.
\begin{equation}
\zeta(s)=  \sum_{n=1}^{\infty}n^{-s}=\prod_{\{p\}}(1-p^{-s})^{-1},\,\,\,\,\,\,
\Re(s)>1,\label{eq:1}
\end{equation}
where $p$ denotes the prime numbers. 

The Riemann zeta function can  be defined in the complex plane
by the contour integral \cite{edwards,titchmarsh,abramowitz,havil}
\begin{equation}
\zeta(s)  =\frac{\Gamma(1-s)}{2\pi i}\int_{C}\frac{t^{s-1}}{e^{-t}-1}dt,
\label{eq:2}
\end{equation}
where the contour of integration $C$ encloses the negative $t$-axis,
looping from $t=-\infty-i0$ to $t=-\infty+i0$ enclosing the point
$t=0$. It is analytic at all points in the complex $s$-plane except
for a simple pole at $s=1$.

The  Riemann zeta function satisfies the following reflection formula:
\begin{equation}
\zeta(s)  =2^{s}\pi^{s-1}\sin\left(\frac{\pi s}{2}\right)\Gamma(1-s)\zeta(1-s),\label{eq:3}
\end{equation}
which shows that the zeta function vanishes when $s$ is a negative
even integer. These zeros are called the trivial zeros of the Riemann
zeta function. All the other zeros of the zeta function are called nontrivial 
zeros. 

The Riemann hypothesis states that all of the nontrivial zeros of
the Riemann zeta function lie on the critical line if  $\Re(s)=1/2$,
i.e.,
\begin{equation}
\zeta\left(\frac{1}{2}+i\lambda_{*}\right)  =0,\label{eq:4}
\end{equation}
where $\lambda_{*}$ is real. The Riemann hypothesis has interesting
connections with different areas of physics and mathematics, such
as quantum mechanics, probability theory, quantum chaos, and quantum
statistical physics \cite{schumayer,wolf}. Therefore, understanding
the Riemann hypothesis from  a supersymmetric approach will have significant
implications in several areas of science. It has been numerically
verified (see, e.g. Ref.~\cite{odlyzko}) that tens of trillions of  nontrivial
zeros of the zeta function exist; they all satisfy the Riemann hypothesis.
To date, various attempts \cite{wolf} have been made to prove the
Riemann hypothesis, but it remains an open problem. 

\section{A supersymmetric model}
\label{sec:model}

%\section{A supersymmetric model}
%\label{sec:3}
We start by defining operators $\Omega$ and $\Omega^{\dagger}$
in terms of the scaling operator $x\frac{d}{dx}$ as
%\begin{subequations}
%\label{eq:1617}
\begin{equation}
\Omega=\frac{\Gamma\left(x\frac{d}{dx}+1\right)}{2\pi i}\intop_{C}
\frac{t^{-x\frac{d}{dx}
-1}}{e^{-t}-1}dt,  \label{eq:16}
\end{equation}
%\Omega^{\dagger}&=&\frac{\Gamma\left(-x\frac{d}{dx}\right)}{2\pi i}\intop_{C}
%\frac{t^{x\frac{d}{dx}}}{e^{-t}-1}dt.\label{eq:17}
%\end{eqnarray}
%\end{subequations}
%Note that we have used the adjoint property
and its adjoint given in terms of the property
\begin{equation}
\left(x\frac{d}{dx}\right)^\dagger=-1-x\frac{d}{dx}.\label{self-adjoint}
\end{equation}
%(See Sec.~\ref{sec:6}.) 
The contour $C$ is as given in Eq.~(\ref{eq:2}).
% the contour of integration encloses the negative $t$-axis, looping
%from $t=-\infty-i0$ to $t=-\infty+i0$ enclosing the point $t=0$.
The operator $\Omega^{\dagger}$ (without a multiplicative factor 
$\Gamma\left(-x\frac{d}{dx}\right)$)
has already been introduced for the study of Riemann zeta functions
from a different perspective \cite{bender1}. The scaling operator $x\frac{d}{dx}$ and its variant have been used as
Hamiltonians \cite{berry2,bender2,bender3} for earlier studies of
Riemann zeros.
In Eq.~(\ref{eq:16}), $\Gamma\left(x\frac{d}{dx}+1\right)$ is defined as
\begin{equation}
\Gamma\left(x\frac{d}{dx}+1\right)=  \int_{0}^{\infty}e^{-y}y^{x\frac{d}{dx}}dy.\label{eq:18}
\end{equation}
%The operators $\Omega$ and $\Omega^{\dagger}$ involve the scaling operator
%$x\frac{d}{dx}$, which counts the powers of $x$ in a monomial according to
%the following, 
%\begin{equation}
%f\left(x\frac{d}{dx}\right)x^{-s}  =f(-s)x^{-s},\label{eq:19}
%\end{equation}
%Equation~(\ref{eq:19}) shows that $x^{-s}$ is an eigenstate of $x\frac{d}{dx}$
%as well as of any operator in the form of $f\left(x\frac{d}{dx}\right)$. 
The operator $\Omega$ and  $\Omega^{\dagger}$ act on the
monomial $x^{-s}$ as 
%\begin{subequations}
%\label{eq:2021}
\begin{equation}
\Omega x^{-s}= \zeta(s)x^{-s},\quad
\Omega^{\dagger}x^{-s}  =\zeta(1-s)x^{-s},\label{eq:21}
\end{equation}
%\end{subequations}
%\begin{equation}
%\Omega x^{-s} %\zeta(s)x^{-s},\label{eq:20}
%\end{equation}
%according to Eq.~(\ref{eq:2}).
% Similarly, we can obtain
%\begin{equation}
%\Omega^{\dagger}x^{-s}  =\zeta(1-s)x^{-s}.\label{eq:21}
%\end{equation}
These operators $\Omega$ and $\Omega^{\dagger}$ have  eigenvalues
in terms of the Riemann zeta function,
and both have the same eigenfunctions.
% Eq.~(\ref{eq:20}) and Eq.~(\ref{eq:21})
%implies that $x^{-s}$ turns out to be the eigenfunctions of both
%the operators $\Omega$ and $\Omega^{\dagger}$, and 
Thus, insofar as these eigenfunctions are complete (which we will address
below),
\begin{equation}
\left[\Omega,\Omega^{\dagger}\right]  =0.\label{eq:22}
\end{equation}
Since the operators $\Omega$ and $\Omega^{\dagger}$ commute, they cannot be 
the ladder (lowering/raising) operators in the supersymmetric
model. Therefore, we introduce a real parameter $\omega\neq0$ and
define the lowering and raising operators of the supersymmetric system
as
%\begin{subequations}
\begin{equation}
A =x^{-i\omega}\Omega,\quad
A^{\dagger}  =\Omega^{\dagger}x{}^{i\omega}.\label{eq:23}
\end{equation}
%\end{subequations}
Now we define a (unnormalized) wavefunction 
\begin{equation}
\phi^{\mu,\rho}(x)  =x^{-\frac{1}{2}+\mu+i\rho},\label{eq:24}
\end{equation}
where $\mu$ and $\rho$ are real parameters. The operators $A$ and
$A^{\dagger}$ %(defined in Eq.~(\ref{eq:23})) 
act on the function $\phi^{\mu,\rho}(x)$ according to 
\begin{subequations}
\label{eq:25c}
\begin{eqnarray}
A\phi^{\mu,\rho}(x)&= & \zeta\left(\frac{1}{2}-\mu-i\rho\right)
\phi^{\mu,\rho-\omega}(x),\\
A^{\dagger}\phi^{\mu,\rho}(x)&= & \zeta\left(\frac{1}{2}+\mu+i(\rho+\omega)
\right)\phi^{\mu,\rho+\omega}(x),\label{eq:25}
\end{eqnarray}
\end{subequations}
where we have used Eq.~(\ref{eq:21}). 
Equation~(\ref{eq:25c}) implies that when the operators $A$ and $A^{\dagger}$
act on the $\phi^{\mu,\rho}(x)$,  the parameter $\rho$
in the wavefunction changes by an amount $\mp\omega$, respectively.

Now,
we construct supersymmetric partner Hamiltonians, $H_{-}$ and $H_{+}$
by
%\begin{subequations}
\begin{equation}
H_{-}  =A^{\dagger}A=\Omega^{\dagger}\Omega,\quad
H_{+}  =AA^{\dagger}=x{}^{-i\omega}\Omega\Omega^{\dagger}x{}^{i\omega}.
\label{eq:26a}
\end{equation}
%\end{subequations}
%When the Hamiltonians $H_{-}$ and $H_{+}$ act on the excited states of the system, it gives the zeta function as an eigenenergy of the system.
%$\phi^{\mu,\rho}(x)$,
%they yield 
%\begin{subequations}
%\begin{eqnarray}
%H_{-}\phi^{\mu,\rho}&= & \zeta\left(\frac{1}{2}-\mu-i\rho\right)
%\zeta\left(\frac{1}{2}+\mu+i\rho\right)\phi^{\mu,\rho},\label{eq:27}\\
%H_{+}\phi^{\mu,\rho}&= & \zeta\left(\frac{1}{2}+\mu+i(\rho+\omega)\right)
%\nonumber\\
%&&\qquad\times\zeta\left(\frac{1}{2}-\mu-i(\rho+\omega)\right)\phi^{\mu,\rho}.
%\label{eq:28}
%\end{eqnarray}
%\end{subequations}
We proceed to discuss the excited states of the supersymmetric
system and show that when supersymmetric partner Hamiltonians  $H_{-}$ and 
$H_{+}$ act on the excited states of the system, the  degenerate eigenenergies
are given in terms of zeta functions.
%\section{Excited state of the supersymmetric model}
%\label{sec:4}
The $n^{th}$ partner states of the supersymmetric system 
$\phi_{n}^{\mu,\rho}(x)$
and $\tilde{\phi}_{n}^{\mu,\rho}(x)$ can be obtained by using 
Eq.~(\ref{eq:25c})
\begin{subequations}
\begin{eqnarray}
\phi_{n}^{\mu,\rho}(x) & =&\frac{A^{\dagger}}{c_{n,-\mu}^{*}}\phi_{n-1}^{\mu,\rho}(x)=x^{-\frac{1}{2}+\mu+i(\rho+n\omega)}, \\
\tilde{\phi}_{n}^{\mu,\rho}(x)&= & A\phi_{n}^{\mu,\rho}(x)=c_{n,\mu}\phi_{n-1}^{\mu,\rho}(x),\label{eq:29}
\end{eqnarray}
\end{subequations}
where
\begin{equation}
c_{n,\mu}=  \zeta\left(\frac{1}{2}-\mu-i(\rho+n\omega)\right).\label{eq:30}
\end{equation}
% Reiterating, the lowering and raising operators
% $A$ and $A^{\dagger}$ act on the $n^{th}$ states of the system
% according to
%\begin{subequations}
%\begin{eqnarray}
%A\phi_{n}^{\mu,\rho}(x) & =& c_{n,\mu}\phi_{n-1}^{\mu,\rho}(x), \\
%A^{\dagger}\phi_{n}^{\mu,\rho}(x) & =&c_{n+1,-\mu}^{*}\phi_{n+1}^{\mu,\rho}(x),
%\label{eq:31}
%\end{eqnarray}
%\end{subequations}
% while $A$ and $A^{\dagger}$ acting on the partner states provide
%\begin{subequations}
% \begin{eqnarray}
% A\tilde{\phi}_{n}^{\mu,\rho}(x) & =&
% c_{n,\mu}\tilde{\phi}_{n-1}^{\mu,\rho}(x),\\
% A^{\dagger}\tilde{\phi}_{n}^{\mu,\rho}(x) & =&
% c_{n,-\mu}^{*}\frac{c_{n,\mu}}{
% c_{n+1,\mu}}
% \tilde{\phi}_{n+1}^{\mu,\rho}(x).\label{eq:32}
% \end{eqnarray}
% \end{subequations}
Then, the Hamiltonians $H_{-}$ acting on $\phi_{n}^{\mu,\rho}(x)$ gives
%and $H_{+}$
%acting on $\tilde{\phi}_{n}^{\mu,\rho}(x)$ give
%\begin{subequations}
\begin{equation}
H_{-}\phi_{n}^{\mu,\rho}(x)  =c_{n,\mu}c_{n,-\mu}^{*}\phi_{n}^{\mu,\rho}(x)
=E_{n,\mu}\phi_{n}^{\mu,\rho}(x), \label{eq:33}
\end{equation}
and $H_+$ has the same eigenvalue acting on $\tilde{\phi}_n^{\mu,\rho}$.
%H_{+}\tilde{\phi}_{n}^{\mu,\rho}(x) & =&c_{n,\mu}c_{n,-\mu}^{*}
%\tilde{\phi}_{n}^{\mu,\rho}(x)=E_{n,\mu}\tilde{\phi}_{n}^{\mu,\rho}(x),
%\label{eq:33}
%\end{eqnarray}
%\end{subequations}
%where
% \begin{eqnarray}
% E_{n,\mu} & =&c_{n,\mu}c_{n,-\mu}^{*}\nonumber \\
%  & =&\zeta\left(\frac{1}{2}-\mu-i(\rho+n\omega)\right)
% \zeta\left(\frac{1}{2}+\mu+i(\rho+n\omega)\right).\nonumber\\
% &&\label{eq:34}
% \end{eqnarray}
%Equation~(\ref{eq:33}) shows that when the supersymmetric partner Hamiltonians
%$H_{-}$ and $H_{+}$ act on the partner states $\phi_{n}^{\mu,\rho}$
%and $\tilde{\phi}_{n}^{\mu,\rho}$, they share the same eigenvalue
The common eigenvalue of $H_\pm$ is
$E_{n,\mu}=c_{n,\mu}c_{n,-\mu}^{*}$, which in general is not real, 
because  $c_{n,\mu}$
and $c_{n,-\mu}^{*}$ are not complex conjugates of each other. We
now address the role of $\mu$.
%can understand this definition of the conjugate by defining the modified
%inner product 
%in the next section.

%\section{Supersymmetric model by using the modified inner product}
%\label{sec:5}
\section{Modified inner product}
\label{sec:mip}
The Dirac adjoint of an operator can be defined in a Hilbert space 
$\mathcal{H}$ with an inner product such that 
\begin{equation}
\langle\psi| H\phi\rangle  =\langle H^{\dagger}\psi|\phi\rangle,\label{eq:35}
\end{equation}
where $|\phi\rangle$,$|\psi\rangle\in\mathcal{H}$. A Hamiltonian
is said to be Hermitian (or self-adjoint) if 
%\begin{align}
$H^{\dagger}  =H$.
%\end{align}
We define an operator $M$ that changes the sign of $\mu$,
%\begin{equation}
%\mu\ensuremath{\stackrel{M}{\longrightarrow}}-\mu,\,\,  M^{\dagger}=M^{-1},
%\label{eq:37}
%\end{equation}
that is,
\begin{equation}
M\phi_{n}^{\mu,\rho}(x)=  \phi_{n}^{-\mu,\rho}(x),\quad M^\dagger=M^{-1}.
\label{eq:38}
\end{equation}
This completely defines $M$ provided these states are complete, which we
establish below. %in Sec.~\ref{sec:7}.
The operator $M$ allows us to define a modified inner product of two
states as 
\begin{equation}
\langle\psi|\phi\rangle_{M}  =\langle\psi|M|\phi\rangle=\langle\psi
| M\phi\rangle=\langle M\psi|\phi\rangle.\label{eq:39}
\end{equation}
In terms of the $M$ inner product, the adjoint is given by 
%Any quantum mechanical system whose Hamiltonians satisfy 
the following similarity transformation 
\begin{equation}
H^{\ddagger}  =M^{-1}H^{\dagger}M;\label{eq:40}
\end{equation}
if this were equal to $H$, this would be the
so-called  pseudo-Hermitian Hamiltonian \cite{pt1,pt2,pt3,pt4,pt5,pt6,pt7}.
In particular, when $M=\mathds{1}$, a pseudo-Hermitian Hamiltonian
coincides with a Hermitian Hamiltonian.  In our case, however, the
Hamiltonians are Hermitian in the usual sense.

We see that the modified inner product defined in 
Eq.~(\ref{eq:39})
brings in a very important feature of the Hilbert space. Namely, when
$\mu\neq0$, the Hilbert space develops a natural modified inner product
different from the Dirac inner product. When $\mu=0$, the modified
Hilbert space coincides with the Hilbert space under the Dirac inner
product.

\section{Boundary conditions, orthonormality, and completeness}
\label{sec:bc}
So far, we have not specified the domain of our wavefunction.  It turns out
that we must define the Hilbert space on a finite interval, which we will
take to be $x\in [1,a]$, $a>1$.  That is, we consider the space of square-integrable
functions
%We define the boundary condition in the domain of the Hilbert space
$L^{2}(1,a)$, with  the periodic  boundary condition
\begin{equation}
\phi_{n}^{\mu,\rho}(1)=  a^{\frac{1}{2}-\mu}\phi_{n}^{\mu,\rho}(a).\label{eq:41}
\end{equation}
This implies
\begin{equation}
1=\exp[  i(\rho+n\omega)\log a].\label{eq:42}
\end{equation}
This condition is satisfied if 
%\begin{subequations}\label{eq:43c}
\begin{equation}
\rho\log a=2\pi k, \quad
\omega\log a=2\pi l\label{eq:43c}
\end{equation}
%\end{subequations}
where $k$ and $l$ are integers.  The simplest case is
 $l=k=1$, which means
\begin{equation}
\rho=\omega=  \frac{2\pi}{\log a}.\label{eq:44}
\end{equation}  %We will consider this case for now, but will return to the
%more general situation later.

We see from Eqs.~(\ref{eq:16}) and (\ref{eq:26a}), that the Hamiltonian
$H_{-}$ can be  written in terms of
the scaling operator  $x\frac{d}{dx}$ as
\begin{eqnarray}
H_{-}&= & -\frac{1}{4\pi^{2}}\Gamma\left(x\frac{d}{dx}+1\right)
\Gamma\left(-x\frac{d}{dx}\right)\nonumber\\
&&\quad\times\intop_C\frac{t^{-x\frac{d}{dx}-1}u^{x\frac{d}{dx}}}
{\left(e^{-t}-1\right)\left(e^{-u}-1\right)}dtdu.\label{eq:45}
\end{eqnarray}
It is easy to  check that the adjoint condition of the operator 
$x\frac{d}{dx}$ (\ref{self-adjoint})
in  the Hilbert space $L^{2}(1,a)$ using the boundary
condition (\ref{eq:41}) implies that $H_-$ is self-adjoint.
%\begin{eqnarray}
%&&\int_{1}^{a}(\phi_{n}^{\mu,\rho})^{*}\left(x\frac{d}{dx}\right)M
%\phi_{n'}^{\mu,\rho}dx 
%\nonumber\\
%&&\quad= \int\phi_{n'}^{\rho,-\mu}\left(-x\frac{d}{dx}-1\right)
%(\phi_{n}^{\rho,\mu})^{*}dx,\label{eq:47}
%\end{eqnarray}
%Using the boundary condition (\ref{eq:41}), we have, therefore,
%\begin{eqnarray}
%&&\int_{1}^{a}(\phi_{n}^{\rho,\mu})^{*}\left(x\frac{d}{dx}\right)
%\phi_{n'}^{\rho,-\mu}dx
%\nonumber\\
%&&\quad= \int\phi_{n'}^{\rho,-\mu}\left(-x\frac{d}{dx}-1\right)
%(\phi_{n}^{\rho,\mu})^{*}dx,\label{eq:47}
%\end{eqnarray}
%which is to say
%\begin{equation}
%\left(x\frac{d}{dx}\right)^{\dagger}=(-x\frac{d}{dx}-1) .\label{eq:48}
%\end{equation}
%as already stated in Eq.~(\ref{self-adjoint}).
%Indeed,  taking the adjoint of  both sides of  Eq.~(\ref{eq:45}),
%and using Eq.~(\ref{eq:48}), we get{\small{}
%\begin{equation}
%H_{-}^{\dagger}=  -\frac{\Gamma(-x\frac{d}{dx})\Gamma(x\frac{d}{dx}+1)}{4\pi^{2}}
%\intop_{C_{1},C_{2}}\frac{t^{x\frac{d}{dx}}u{}^{-x\frac{d}{dx}-1}}{\left(e^{-t}-1\right)\left(e^{-u}-1\right)}dtdu;\label{eq:49}
%\end{eq uation}
%}Now interchanging the dummy variables in the integration, we get
%\begin{equation}
%H_{-}^{\dagger}=  H_{-}.\label{eq:50}
%\end{equation}

%\section{Orthonormality and completeness}
%\label{sec:7}
%\subsection{Normalization condition}
\label{sec:7a}
We define the Hilbert space in the domain $L^{2}(1,a)$ and show that
$\phi_{n}^{\mu,\rho}(x)$ forms an orthonormal set of functions as
\begin{eqnarray}
\langle n,\mu| n',\mu\rangle_{M}&= & \langle n,\mu| n',-\mu\rangle
=  \int_{1}^{a}(\phi_{n}^{\mu,\rho})^{*}\phi_{n'}^{-\mu,\rho}dx\nonumber \\
&= & \int_{1}^{a}x^{-1+i(n-n')\omega}dx
=  \delta_{nn'}\log a,\label{eq:51}
\end{eqnarray}
where $\phi_{n}^{\mu,\rho}(x)=\langle x| n,\mu\rangle$. 
We see that there is an extra factor of $\log a$ multiplying the Kronecker 
delta function $\delta_{nn'}$.
Therefore, we define the wave function as
\begin{equation}
\psi_{n}^{\mu,\rho}(x)  =\frac{1}{\sqrt{\log a}}\phi_{n}^{\mu,\rho}(x)
  =\frac{1}{\sqrt{\log a}}x^{-\frac{1}{2}+\mu+i(\rho+n\omega)}.\label{eq:52}
\end{equation}
which is orthonormal for all values of $k$ and $l$.

%\subsection{Completeness}
%\label{sec:7b}
We can establish the completeness condition by considering
\begin{eqnarray}
&&\sum_{n=-\infty}^{\infty}\psi_{n}^{\mu,\rho}(x)M\left(\psi_{n}^{\mu,\rho}(y)
\right)^{*}  =\sum_{n=-\infty}^{\infty}\psi_{n}^{\mu,\rho}(x)
\left(\psi_{n}^{-\mu,\rho}(y)\right)^{*}\nonumber \\
&&\qquad=  \frac{1}{\log a}\frac{1}{y}
\left(\frac{y}{x}\right)^{\frac{1}{2}-\mu-i\rho}
\sum_{n=-\infty}^{\infty}e^{in\omega(\log x-\log y)}.\label{eq:53}
\end{eqnarray}
Now because $-\log a<\log x-\log y<  \log a$, and  using Eq.~(\ref{eq:44}) for
$\omega$, 
 the Poisson summation formula implies that
we can write Eq. (\ref{eq:53}) as
%\begin{equation}
%\sum_{n=-\infty}^{\infty}e^{in(x-y)}=  %\sum_{k=-\infty}^{\infty}2\pi\delta(x-y+2k\pi),\label{eq:54}
%\end{equation}
%and restrict 
%\begin{equation}
%-\pi<\log x-\log y<  \pi,\label{eq:55}
%\end{equation}
%we can write Eq.~(\ref{eq:53}) as
%\begin{eqnarray}
%&&\sum_{n=-\infty}^{\infty}\psi_{n}^{\mu,\rho}(x)\left(\psi_{n}^{-\mu,\rho}(y)
%\right)^{*} \nonumber\\
%&&\qquad =\frac{2\pi}{\log a}\frac{1}{y}(\frac{y}{x})^{\frac{1}{2}-\mu-i\rho}
%\delta(\omega\log(\frac{x}{y})).\label{eq:56}
%\end{eqnarray}
%Using the property of the Dirac delta function,
%\begin{equation}
%\delta(f(x))=  \frac{1}{f'(x)}\delta(x),\label{eq:57}
%\end{equation}
%we can write Eq.~(\ref{eq:56}) as
\begin{eqnarray}
\sum_{n=-\infty}^{\infty}\psi_{n}^{\mu,\rho}(x)\left(\psi_{n}^{-\mu,\rho}(y)\right)^{*}&=\delta(x-y).\label{eq:58}
\end{eqnarray}
%In the last line, we have used the property of the Dirac delta function
%\begin{align}
%f(x)\delta(x-a)= & f(a)\delta(x-a).\label{eq:59}
%\end{align}
%again using Eq.~(\ref{eq:44}). %we can write Eq.~(\ref{eq:58}) as
%the desired completeness statement,
%\begin{eqnarray}
%\sum_{n=-\infty}^{\infty}\psi_{n}^{\mu,\rho}(x)\left(\psi_{n}^{-\mu,\rho}(y)\rig%ht)^{*}= & \delta(x-y).\label{eq:60}
%\end{eqnarray}
The completeness relation only holds for $l=1$, but imposes no restriction
on $k$.
Equations (\ref{eq:51}) and  (\ref{eq:58}) show that under the modified
inner product the functions $\psi_{n}^{\mu,\rho}(x)$ form a complete, 
 orthonormal set of functions, provided $\omega\ln a=2\pi$.

\section{Degeneracies}
\label{sec:8}
In our model, we see that the states $\tilde{\psi}_{n}^{\mu,\rho}(x)=c_{n,\mu}\psi_{n-1}^{\mu,\rho}(x)$
and $M\left(\psi_{n+1}^{\mu,\rho}(x)\right)^{*}=\left(\psi_{n+1}^{-\mu,\rho}(x)\right)^{\star}$
have the same energies when they are acted upon  by $H_{+}$, i.e.,
\begin{subequations}
\begin{eqnarray}
H_{+}\tilde{\psi}_{n}^{\mu,\rho}(x)&= & E_{n,\mu}\tilde{\psi}_{n}^{\mu,\rho}(x),
\\
H_{+}\left(\psi_{n+1}^{-\mu,\rho}(x)\right)^{\star}&= & E_{n,\mu}\left(\psi_{n+1}^{-\mu,\rho}(x)\right)^{\star}.\label{eq:61}
\end{eqnarray}
\end{subequations}
Thus, the states $\phi_{n-1}^{\mu,\rho}$ and $\left(\phi_{n+1}^{-\mu,\rho}
\right)^*$ are
degenerate under $H_+$.  Even more obviously, $\phi_n^{\mu,\rho}$ and 
$\left(\phi_n^{-\mu,\rho}\right)^*$ are degenerate under $H_-$, since 
these two states are given by $x^{-1/2\pm\mu\pm i(\rho+n\omega)}$. Further
aspects of degeneracy will be explored in the next section.

\section{Reality and spectrum}
\label{sec:rs}
Under the Dirac inner product,
a Hermitian Hamiltonian is guaranteed to have real eigenvalues.  What
happens in the case of the $M$ inner product?
We first consider the matrix element of the  Hamiltonian $H_{-}$ under
the modified inner product as follows:
%\begin{eqnarray}
%\langle\psi_{n'}^{\mu,\rho}\mid H_{-}\psi_{n}^{\mu,\rho}\rangle_{M} & =&
%\langle\psi_{n'}^{\mu,\rho}\mid MH_{-}\psi_{n}^{\mu,\rho}\rangle\nonumber \\
% & =&\int_{1}^{a}dx\,\left(\psi_{n'}^{\mu,\rho}\right)^*E_{n,\mu}M
%\psi_{n}^{\mu,\rho}\nonumber \\
% & =&\int_{1}^{a}dx\,\left(\psi_{n'}^{\mu,\rho}\right)^{*}E_{n,\mu}
%\psi_{n}^{-\mu,\rho}\nonumber \\
% & =&E_{n,\mu}\delta_{n,n'},\label{eq:62}
%\end{eqnarray}
\begin{equation}
\langle\psi_{n'}^{\mu,\rho}| H_{-}\psi_{n}^{\mu,\rho}\rangle_{M}  
=E_{n,\mu}\delta_{n,n'},\label{eq:62}
\end{equation}
and then the corresponding element for the adjoint, given in Eq.~(\ref{eq:40}):
%\begin{eqnarray}
%\langle H_{-}^{\ddagger}\psi_{n'}^{\mu,\rho}\mid\psi_{n}^{\mu,\rho}
%\rangle_{M}&= & \langle H_{-}^{\dagger}M\psi_{n'}^{\mu,\rho}\mid
%\psi_{n'}^{\mu,\rho}\rangle\nonumber \\
%&= & \int_{1}^{a}dx\left(H_{-}M(\psi_{n'}^{\mu,\rho})^{*}\right)
%\psi_{n}^{\mu,\rho}\nonumber \\
%&= & \int_{1}^{a}dx\left(H_{-}(\psi_{n'}^{-\mu,\rho})^{*}\right)
%\psi_{n}^{\mu,\rho}\nonumber \\
%&= & E_{n,-\mu}^{*}\delta_{n,n'}.\label{eq:63}
%\end{eqnarray}
\begin{equation}
\langle H_{-}^{\ddagger}\psi_{n'}^{\mu,\rho}|\psi_{n}^{\mu,\rho}
\rangle_{M}=  E_{n,-\mu}^{*}\delta_{n,n'}.\label{eq:63}
\end{equation}
%we can write Eq. (\ref{eq:63}) as
%\begin{equation}
%\langle H_{-}^{\ddagger}\psi_{n'}^{\mu,\rho}\mid\psi_{n}^{\mu,\rho}\rangle_{M} 
% =E_{n,-\mu}^{*}\delta_{n,n'}\label{eq:64}
%\end{equation}
%where $H_{-}^{\ddagger}=M^{-1}H_{-}^{\dagger}M$.
From Eqs.~(\ref{eq:62}) and (\ref{eq:63}), we see that 
\begin{equation}
E_{n,\mu}=E^*_{n,-\mu};
\end{equation}
when $\mu=0$ all the energies are real, but this is not true if $\mu\ne0$.
In the latter case, only for a special situation 
can even the ground-state energy be real, as we will now see.
%it gives real
%and positive ground state energy only in two cases. We study the ground
%state of the system for these two cases in the next section.

%\section{Nontrivial and Trivial zeros of the Riemann zeta function}
%\label{sec:10}
%\subsection{Case 1: Ground state energy when $\mu=0$ and $\rho=\omega$}
%\label{sec:10a}
The simplest possibility occurs when $\rho=\omega$, that is, $k=1$ as well
as $l=1$.  Then,
for $\mu=0$, the  Hamiltonian
acting on the $n^{th}$ state of the system gives
\begin{equation}
H_{-}\psi_{n}^{0,\omega}(x)  =\left|\zeta\left(\frac{1}{2}+i(n+1)\omega\right)
\right|^{2}\psi_{n}^{0,\omega}(x).\label{eq:65}
\end{equation}
%where
%\begin{equation}
%E_{n,0}  =\zeta\left(\frac{1}{2}-i(n+1)\omega\right)\zeta\left(
%\frac{1}{2}+i(n+1)\omega\right).\label{eq:66}
%\end{equation}
In this case, the modified inner product [Eq.~(\ref{eq:39})]
becomes  the Dirac inner product [Eq.~(\ref{eq:35})]
and the ground state of the supersymmetric system vanishes
%\begin{equation}
%H_{-}\psi_{0}^{0,\omega}(x)  =\left|\zeta\left(\frac{1}{2}+i\omega\right)
%\right|^{2}
%\psi_{0}^{0,\omega}(x).\label{eq:67}
%\end{equation}
%We can make this energy vanish, as required for  the supersymmetric system,
provided we choose $\omega$ to correspond to one of the nontrivial zeros of 
the zeta function, $\omega=\lambda_*$, as given by Eq.~(\ref{eq:4}).
%\begin{equation}
%\zeta\left(\frac{1}{2}+i\lambda_*\right)  =0,\label{eq:68}
%\end{equation}
%where we recall that the nontrivial zeros of the zeta function lie along
%the line $\Re (s)=1/2$, according to the Riemann hypothesis.
%which shows that zeros of the zeta function occur when $\Re(s)=1/2$,
%which is the }\textit{\textcolor{black}{Riemann hypothesis}}\textcolor{black}{.
%It means, }in Eq.~(\ref{eq:68}), $\omega$ corresponds to the location
%of non-trivial zeros of the Riemann zeta function. \\

%\subsection{Case 2: Ground state energy when $\rho=0$}
%\label{sec:10b}
On the other hand, for $\mu\ne0$ and
  $\rho=0$, $H_{-}$ acting on the $n^{th}$ state gives
\begin{equation}
H_{-}\psi_{n}^{\mu,0}(x)=  \zeta\left(\frac{1}{2}-\mu-in\omega\right)
\zeta\left(\frac{1}{2}+\mu+in\omega\right)\psi_{n}^{\mu,0}(x),\label{eq:69}
\end{equation}
where the eigenvalue is almost always complex.
However, for the ground state $n=0$, the eigenvalue is real,
%\begin{equation}
%H_{-}\psi_{0}^{\mu,0}(x)  =\zeta\left(\frac{1}{2}-\mu\right)
%\zeta\left(\frac{1}{2}+\mu\right)\psi_{0}^{\mu,0}(x),\label{eq:70}
%\end{equation}
and can be made equal to zero only if $\frac{1}{2}\pm\mu=-2m$, where $m\in N$. 
This case gives a vanishing  ground state energy for the system,
 corresponding to the trivial zeros of the Riemann zeta function.
However, the excited states all have complex energies, 
so it is still an open problem to understand the physical interpretation 
of this. 
%to call this situation supersymmetric.

\section{Natural supersymmetry}
\label{sec:XI}
Let us
 return to the conditions (\ref{eq:43c}) for the nontrivial zeros.
 In the above, we assumed $k=l=1$,
as the simplest possibility. However, only $l=1$ is required for
completeness;  the general eigenvalues for the 
$\mu=0$ scenario are
\begin{eqnarray}
E_n^{k,i}&=&
\zeta\left(\frac12-i\lambda_*^{(i)}\left(1+\frac{n}k \right)\right)\nonumber\\
&&\quad\times\zeta\left(\frac12+i\lambda_*^{(i)}
\left(1+\frac{n}k \right)\right),\label{eq:nss}
\end{eqnarray}
where the index $i$ denotes the $i$th positive value of $\lambda_*$.
%provided $\lambda_*^{(i)}>2k$,  and $k$ and $l$ are
%given by Eq.~(\ref{eq:43c}).  
In general, these eigenvalues are all real and
positive, except for the zero eigenvalues at $n=0$.  But  
$2k$ is an even integer, so  all eigenvalues are doubled, 
including that for  the $n=0$ state, which has a partner zero energy state for
$n=-2k$.  Even $H_-$ exhibits a kind of ``natural 
supersymmetry.''
There is one exception to this: there is an odd 
number
of states between these two zero-energy states, and thus
there is one  isolated state at $n=-k$ with
positive energy $\zeta(1/2)^2$, which is the only nondegenerate state. 
%(In contrast,
%if $2k/l$ is not an integer, there is no degeneracy and no natural
%supersymmetry.)
 The spectrum is extremely
oscillatory, apparently chaotic, with small eigenvalues appearing whenever the
integer $n$ happens to yield an approximate
 coincidence with another nontrivial zero,
which can never occur exactly.

\begin{figure}
\includegraphics{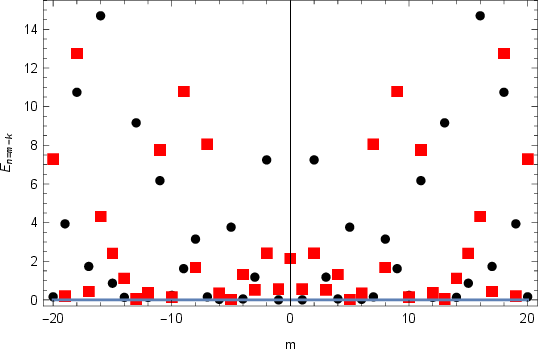}
\caption{\label{fig}
The first 40 eigenenergies $E_{n=m-k}^{k,i}$
of the  model given in Eq.~(\ref{eq:nss}) 
for two cases: 
(1) $i=1$, $k=l=1$, denoted by black dots.  The zero energy states occur
only at $n=0$ and $n=-2$, and the other apparently vanishing energies are 
actually positive.  Note that the spectrum is symmetric around $n=-1$, 
which is the only nondegenerate state.
(2) $i=3$, $k=5$, $l=1$ denoted by  red squares.  The symmetry of the spectrum
is now about $n=-5$.  The only zero energy
states are for $n=0$ and $n=-10$. %and the central point occurs for $m=-5$.
%(3) Shown for comparison are the eigenenergies for $i=3$, $l/k=3/2$,
%denoted by the open circles,  which
%has no degeneracy, but still has zero energy for $n=0$, but nowhere else.
}
\end{figure}

We illustrate these remarks in Fig.~\ref{fig}.
So it appears in that in all realistic situations, supersymmetry
of a sort emerges without the need for a partner Hamiltonian.  

\section{Conclusion}
\label{sec:concl}
In summary, we have defined a supersymmetric quantum mechanical model
whose vanishing ground state energy is consistent with the Riemann zeta 
function $\zeta(s)$ having zeros along the line $\Re(s)=1/2$, 
and also when $s$ is a negative even integer. 
These two cases correspond to the existence of nontrivial
and trivial zeros of the Riemann zeta function, respectively. 
The former case is quite interesting, since  the real spectrum of the
Hamiltonian $H_-$ is supersymmetric, in that all states, including the
ground state $n=0$, are doubly degenerate, with the exception of the state
at $n=-k$, without the need for a partner Hamiltonian.  
Although the spectrum corresponding to the nontrivial zeros is entirely
real, it exhibits a chaotic oscillatory behavior.
On the other hand,
the trivial zero ground state has only complex excited states, so the 
physical significance is obscure.

Our investigations here continue 
our attempt to  understand the connection between
the condition that the ground state energy of the supersymmetric model
vanish, and the location of the zeros of the Riemann zeta function.
While the observations in this paper in no sense constitute a proof of
the Riemann hypothesis, they do lend further credence to it.
Any nontrivial zero not lying on the critical line could not correspond
to a complete set of eigenstates with real energies in our model.
Our approach bears some superficial resemblance to the Hilbert-P\'{o}lya
conjecture \cite{mont}, with the virtue that our Hamiltonian is explicit,
defined in a Hilbert space.

\medskip

%trivial and nontrivial
%zeros corresponding to the vanishing ground state energy of the supersymmetric
%model.
\begin{acknowledgments}
PK would like to thank Prof.~Ashok Das, who has been a wonderful mentor
and great collaborator; his suggestions and feedback on this manuscript
are highly acknowledged. PK  would also like to acknowledge Dr.~Ashok Kumar
Diktiya and Prof.~S. Murugesh 
for many insightful comments on the manuscript.
PK acknowledges support by the DST, Govt.~of India under the
Women Scientist A, Ref.~No.~DST/WOS-A/PM-64/2019 scheme.
The work of KAM is supported in part by a grant from the US
National Science Foundation, grant number  2008417.
\end{acknowledgments}

\end{document}